\documentclass{emulateapj}
\usepackage{lineno}
\pdfoutput=1

\def\ga{\mathrel{\hbox{\rlap{\hbox{\lower4pt\hbox{$\sim$}}}\hbox{$>$}}}}
\def\la{\mathrel{\hbox{\rlap{\hbox{\lower4pt\hbox{$\sim$}}}\hbox{$<$}}}}

\usepackage{verbatim}
\usepackage{amsmath}
\usepackage{multirow}
\usepackage{relsize}
\usepackage{natbib}
\usepackage{color}
\usepackage{placeins}
\usepackage[utf8]{inputenc}
\usepackage[english]{babel}
\usepackage{appendix}
\usepackage[normalem]{ulem}
		
\usepackage{amssymb}
\usepackage{float}

\usepackage{amsmath,amstext,amssymb}
\usepackage{float}
\usepackage{multirow}
\usepackage{relsize}
\usepackage{natbib}

\usepackage[dvipsnames]{xcolor}
\usepackage{hyperref}
\usepackage{cleveref}

\newcommand\myshade{85}
\colorlet{mylinkcolor}{violet}
\colorlet{mycitecolor}{Blue}
\colorlet{myurlcolor}{Aquamarine}

\hypersetup{
  linkcolor  = mylinkcolor!\myshade!black,
  citecolor  = mycitecolor!\myshade!black,
  urlcolor   = myurlcolor!\myshade!black,
  colorlinks = true,
}

\DeclareSymbolFont{matha}{OML}{txmi1}{m}{it}
\DeclareMathSymbol{\varv}{\mathord}{matha}{118}

\def\ga{\mathrel{\hbox{\rlap{\hbox{\lower4pt\hbox{$\sim$}}}\hbox{$>$}}}}
\def\la{\mathrel{\hbox{\rlap{\hbox{\lower4pt\hbox{$\sim$}}}\hbox{$<$}}}}

\usepackage{booktabs}
\usepackage{url}
\usepackage{scalerel,tikz}
\usetikzlibrary{svg.path}
\definecolor{orcidlogocol}{HTML}{A6CE39}
\tikzset{orcidlogo/.pic={
 \fill[orcidlogocol] svg{M256,128c0,70.7-57.3,128-128,128C57.3,256,0,198.7,0,128C0,57.3,57.3,0,128,0C198.7,0,256,57.3,256,128z};
 \fill[white] svg{M86.3,186.2H70.9V79.1h15.4v48.4V186.2z}
 svg{M108.9,79.1h41.6c39.6,0,57,28.3,57,53.6c0,27.5-21.5,53.6-56.8,53.6h-41.8V79.1z M124.3,172.4h24.5c34.9,0,42.9-26.5,42.9-39.7c0-21.5-13.7-39.7-43.7-39.7h-23.7V172.4z}
 svg{M88.7,56.8c0,5.5-4.5,10.1-10.1,10.1c-5.6,0-10.1-4.6-10.1-10.1c0-5.6,4.5-10.1,10.1-10.1C84.2,46.7,88.7,51.3,88.7,56.8z};
}}
\newcommand\orcidicon[1]{\href{https://orcid.org/#1}{\mbox{\scalerel*{
\begin{tikzpicture}[yscale=-1,transform shape]
\pic{orcidlogo};
\end{tikzpicture}
}{|}}}}

\usepackage{ulem}

\shorttitle{Interstellar Driving Scales Revealed by Bispectrum}

\shortauthors{O'Brien, Burkhart \& Shelley}
\begin{document}
\title{Studying Interstellar Turbulence Driving Scales using the Bispectrum}

\author{Michael J. O'Brien, \altaffilmark{1, 2}~\orcidicon{ } }
\altaffiltext{1}{Flatiron Institute, 162 Fifth Avenue, New York, NY 10010, USA}
\altaffiltext{2}{Department of Physics, Harvard University, 17 Oxford St, Cambridge, MA 02138, USA}

\author{Blakesley Burkhart, \altaffilmark{1, 3}~\orcidicon{0000-0001-5817-5944} }
\altaffiltext{3}{Department of Physics and Astronomy, Rutgers University,  136 Frelinghuysen Rd, Piscataway, NJ 08854, USA}

\author{Michael J. Shelley, \altaffilmark{1, 4}~\orcidicon{ } }
\altaffiltext{4}{Courant Institute, New York University, New York, NY 10012, USA}

\begin{abstract}
We demonstrate the utility of the bispectrum, the Fourier three-point correlation function, for studying driving scales of magnetohydrodynamic (MHD) turbulence in the interstellar medium. We calculate the bispectrum by implementing a parallelized Monte Carlo direct measurement method, which we have made publicly available. In previous works, the bispectrum has been used to identify non-linear scaling correlations and break degeneracies in lower-order statistics like the power spectrum. We find that the bicoherence, a related statistic which measures phase coupling of Fourier modes, identifies turbulence driving scales using density and column density fields. In particular, it shows that the driving scale is phase-coupled to scales present in the turbulent cascade. We also find that the presence of an ordered magnetic field at large-scales enhances phase coupling as compared to a pure hydrodynamic case.  We therefore suggest the bispectrum and bicoherence as tools for searching for non-locality for wave interactions in MHD turbulence.  
\end{abstract}

\section{Introduction}

The gas and dust that comprises the interstellar medium (ISM) of galaxies is known to be magnetized and in a highly turbulent dynamical state \citep{GS95,ElmegreenScalo,Lazarian07rev,krumreview2014}.
Magnetohydrodynamic (MHD) turbulence in the ISM plays an important role in the evolution of galaxies, including regulating the motion of cosmic rays \citep{Schlickeiser02,LY14,Xu2016b}, mediating the destruction and creation of magnetic fields \citep{LV99,Sant10}, and playing a role in star formation \citep{Lars81,Mckee_Ostriker2007,maclow04,ElmegreenScalo,Collins12,Burkhart2015ApJ...811L..28B,padoan2017ApJ...840...48P,Burkhart2017ApJ...834L...1B,BurkhartMocz2019}. 

To maintain turbulence in a fluid, energy must be injected and replenished at least on a eddy turn over time, $t_{\rm eddy}=L/\sigma_z$, where $L$ is the injection scale of the turbulence and $\sigma_z$ is the injection velocity scale.   After this time, the turbulent eddies have cascaded down to the dissipation scale, where the energy is radiated, converted into heat, or transferred via friction of ions and neutrals. 
There are many possible energy sources for ISM turbulence that may act on a range of scales and in different ISM phases.
These injection sources can include gravitational instabilities and mass accretion onto galaxies \citep{Forbes14a,Goldbaum15a,2016MNRAS.458.1671K,krumholz2018}, protostellar jets \citep{Cunningham_2006,Banerjee_2007,2009ApJ...692..816C,Federrath2015a}, the Magnetorotational Instability \citep{2004A&A...423L..29D}, expansion of H II regions \citep{maclow04}, stellar winds \citep{2015ApJ...811..146O,Gallegos-Garcia:2020:ApJL}, and supernovae explosions \citep{2009MNRAS.394..157L,2011IAUS..274..348G,padoan2017ApJ...840...48P}. The dominant largest-scale injection processes are typically thought to be supernova and disk instabilities, which can act on 100 parsec to kiloparsec scales and would imply a dissipation timescale of $\approx 10-100$ Myrs for Milky Way-like galaxies.  While this time-scale is short, it is longer than or equal to other relevant time scales such as the average supernova rate ($\approx 5-10$ Myrs) or the orbital time ($\approx 100$ Myr) \citep{2016ApJ...822...11P, 10.1093/mnras/stv562,krumholz2018}.  Thus it is likely that a number of injection mechanisms can maintain the observed levels of non-thermal velocity dispersion seen in different ISM tracers however which mechanism dominates is still an open question \citep{Utomo_2019}.

Turbulent motions are self-similar in the inertial range and give rise to the power-law behavior seen in the power spectra, while at the injection scale and dissipation scales, non-power law shapes are observed.
In particular, the dissipation scale shows exponential decay in the turbulent kinetic energy power spectra while the injection scale shows a marked bump at larger scales, followed by a transition to a power-law. 
Additionally, the slope of the power-law depends on a number of physical factors, such as the compressibility of the turbulence, the presence of gravity, thermal instability, and the strength of the magnetic field:
\citep{lp06,padoan97,Kritsuk07a,lazarian09rev,Kowal07,Collins12,burkhart13,2018ApJ...856..136P,Gallegos-Garcia:2020:ApJL}.

In practice, it is difficult to observationally directly measure the kinetic energy power spectrum and the injection scale and several methods have been proposed in the literature. 
When dealing with radio position-position-velocity (PPV) data, the Velocity Coordinate Spectrum (VCS) has been used to directly measure the injection scale of the turbulence \citep{lp06,chepurnov15}. 
\citet{Haverkorn2008} observed Faraday rotation of extragalatic radio sources through the Galactic plane and
suggested that stellar winds and protostellar outflows drive turbulence on
parsec scales in the spiral arms while supernova do the driving at 100 parsec scales in the interarm regions. \cite{Yoon2019ApJ...880..137Y} demonstrated the use of polarization and the Davis-Chandrasekhar-Fermi Method to determine the driving scale of turbulence.
More recently, \citet{bialy2017ApJ...843...92B} and \citet{bialy2020} suggest using chemical transitions as diagnostics of the injection scale, since the injection scale can influence the formation of density fluctuations in the ISM and thereby alter the chemistry. 
Finally, comparison of analytic scaling predictions for different driving mechanisms  can also be useful to shed light on which driving mechanism might be dominant, e.g. the study by \citep{krumholz2018} suggests investigating the star formation rate vs. gas velocity dispersion correlation can discern between supernova vs. mass transport driving mechanisms. 

In this paper we examine the use of higher-order statistics to identify the driving scale of ISM turbulence. In particular, injection of energy can produce correlated Fourier modes and thus techniques that examine mode-mode coupling and preserve phase information may be valuable. 
A commonly-used higher-order signal processing technique that contains information from both Fourier amplitude and phases is the bispectrum \citep{75086}.
The Fourier transform of the second-order cumulant (the autocorrelation function) is the power
spectrum (see Eq. \ref{eq:ps}), while the Fourier transform of the third-order cumulant is known as the bispectrum.
The bispectrum and related three-point correlation function has been used in cosmology to identify key scales of interest in cosmic structure, e.g. Baryon Acoustic Osculations \citep{2007ApJ...664..675E,Slepian15_alg,Slepian16_alg_WFTs,Pearson_2018}.  The bispectrum and three-point correlation function have been applied in the context of turbulence in order to determine key parameters of the flow such as the sonic and Alfv\'enic Mach number \citep{Burkhart09a,Portillo2018ApJ...862..119P,2021ApJ...910..122S}.  However, given its past use in cosmology as a indicator of key physical scales of interest, the bispectrum may also be of interest in galaxy physics for locating turbulent driving scales and scales of filament structure.  The motivation of this work is to investigate the utility of the bispectrum and related higher-order statistics for locating the driving scales of interstellar turbulence.
This paper is organized as follows: in Section \ref{sec:sims} we outline the simulation datasets used with varying driving scale, in Section \ref{sec:bi} we define the bispectrum and bicoherence for our convention, in Section \ref{sec:res}
we show our results of the bispectrum with different turbulence driving scales, and finally in Section \ref{sec:dis} we discuss the implications of our results followed by our conclusions in Section \ref{sec:con}.
Last, the Appendices \ref{appendix:estimator} to \ref{appendix:fftestimator} describe the technical details of our bispectrum implementation.

\begin{figure*}
    \center
    \includegraphics[width=\textwidth]{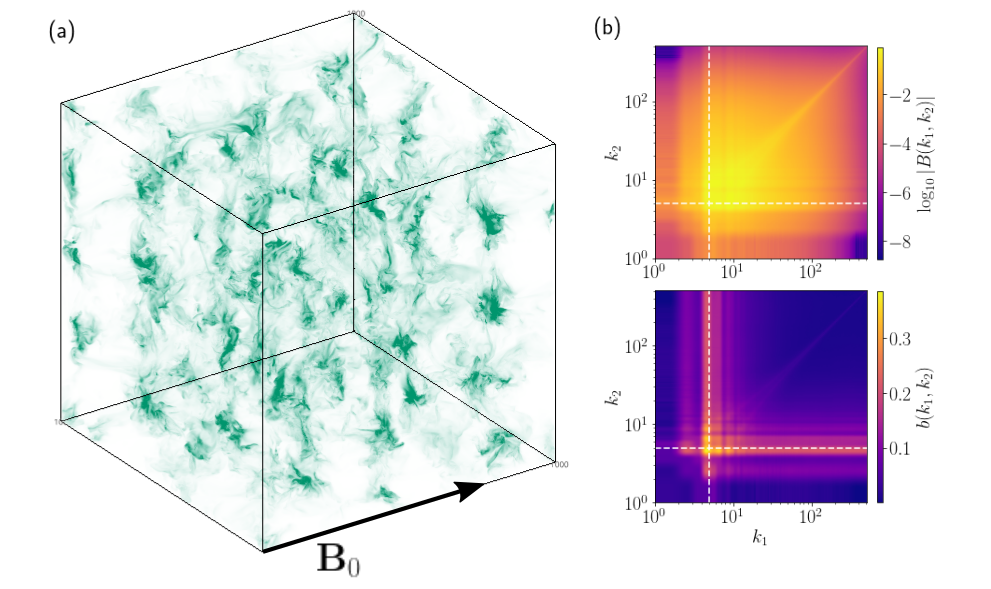}
    \caption{
     (a) Volume rendering for the density field of $N_{\rm res} = 1024^3$ resolution elements and driving scale $k_{\rm drive} = 5$ with sonic Mach number $\mathcal{M}_s = 4.5$ and Alfv\'enic Mach number $\mathcal{M}_A = 0.7$. Volume rendering is done with the open source visualization software Glue.
     (b) The bispectrum and bicoherence of (a). These statistics are able to pick out the turbulence driving scale by identifying phase coupling produced by the Fourier driving mechanism. This is indicated by the dashed black line overlayed onto the L-shaped spikes in the bispectrum and bicoherence. The values of $0.2-0.3$ in the bicoherence can be interpreted as the strength of phase coupling \citep{BiphaseEnergyTransfer}.}
     \label{fig:MainFigure}
\end{figure*}

\section{Simulations}
\label{sec:sims}

We run 3D numerical simulations of
isothermal compressible MHD turbulence with a setup similar to that of a number of past works  \citep{Hill2008,Burkhart09a,Herron2016ApJ...822...13H,bialy2017ApJ...843...92B,bialy2020}.
We refer to these works for the
details of the numerical set-up and here provide a short
overview.  

The code is a third-order accurate
 essentially non-oscillatory method (ENO) 
 scheme which solves the ideal MHD equations
in a periodic box:
\begin{align}
 \frac{\partial \rho}{\partial t} + \nabla \cdot (\rho \pmb{\varv}) &= 0, \\
 \frac{\partial \rho \pmb{\varv}}{\partial t} + \nabla \cdot \left[ \rho \pmb{\varv} \pmb{\varv} + \left( p + \frac{B^2}{8 \pi} \right) {\bf I} - \frac{1}{4 \pi}{\bf B}{\bf B} \right] &= {\bf f},  \\
 \frac{\partial {\bf B}}{\partial t} - \nabla \times (\pmb{\varv} \times{\bf B}) &= 0. \ 
 \label{eq:mhd}
\end{align}
Here ${\bf B}$ is the magnetic field, $p$ is the gas pressure, and ${\bf I}$ is the identity matrix. These simulations have periodic boundary conditions and an isothermal equation of state $p = c_{\rm s}^2 \rho$, with $c_{\rm s}$ the isothermal sound speed.
For the energy source term $\bf{f}$, we assume a random large-scale driving at a wave number which is varied as described below.  The driving is continuous in space and time and constrained to be solenoidal. 
 The models are run for $t \approx 5$ crossing times,
to guarantee full development of the energy cascade.

The code outputs a velocity field $\mathbf{v}$, density field $\rho$, and magnetic field $\mathbf{B}$.
The magnetic field consists of the uniform background
field and a turbulent field, i.e: ${\bf B} = {\bf B}_0 + {\bf b}$ with the magnetic field initialized along a single preferred direction. 

We run simulations with a sonic Mach number $\mathcal{M}_s = 4.5$.
While previous studies used driving on large scales, with $k_{\rm drive}=2-2.5$, here
we run several simulations, each of which is driven on a different peak driving scale, of $k_{\rm drive}=2.5, \ 5, \ 7, \ 10,$ and $20$, with the full driving range about $\Delta k= \pm 1$ around the peak. The sonic Mach number is set to $\mathcal{M}_s = 4.5$ and the Alfv\'en Mach number is $\mathcal{M}_A= 0.7$ for all simulations.
To test numerical convergence we run simulations with various resolutions; $N_{\rm res}=256^3, 512^3$ and $1024^3$ resolution elements.

The simulations presented in this work are entirely in adimensional code units and can be rescaled to different physical scalings.  For a detailed discussion on how to convert code units into physical units, see
Appendix A of \citep{Hill2008} and the discussion in \citep{McKee10b}.

A subset of these simulations are part of the Catalog for Astrophysical Turbulence Simulations (CATS\footnote{See \href{www.mhdturbulence.com}{www.mhdturbulence.com}.}, Burkhart et al. 2020). An example simulation cube is shown in Fig. \ref{fig:MainFigure}.

\section{The Bispectrum and Bicoherence}
\label{sec:bi}

In order to measure the relevant scales of sources and sinks of turbulent energy, it is common to employ correlation functions as statistical descriptors. The Fourier space two-point correlation function is the power spectrum.
Under the assumption of translational invariance, we define the density power spectrum as
\begin{equation}
    P(\mathbf{k}) = \langle |\tilde{\rho}(\mathbf{k})|^2 \rangle,
    \label{eq:2ptcorr}
\end{equation}
\noindent
where $\mathbf{k}$ is a wave vector, $\langle \cdot \rangle$ is an ensemble average, and $\tilde{\rho}(\mathbf{k})$ is the Fourier transform of the density field $\rho(\mathbf{x})$ \citep{bernardeau2002large}.
We restrict our study to the density, but we can in general define power spectra for the velocity and magnetic field vectors or the kinetic and magnetic energy scalars.
Assuming that the system size is large enough that spatial averaging is sufficient as an ensemble average, we compute the isotropic power spectrum
\begin{equation}
P(k)= \sum_{|\mathbf{k}| \in [k, k+\Delta k)}|\tilde{\rho}(\mathbf{k})|^2,
\label{eq:ps}
\end{equation}
where $k$ is the wavenumber and $\Delta k$ is a bin width. In this work we set $\Delta k = 1$. $P(k)$ is often normalized by the bin volume for a particular $k$, however we use the fluid mechanics convention for computing energy spectra.

We also consider the Fourier three-point correlation function, known as the bispectrum. Under the same assumptions as Eq. \ref{eq:2ptcorr} we define the bispectrum as
\begin{align}
\begin{split}
    B(\mathbf{k}_1, \mathbf{k}_2) &= \langle \tilde{\rho}(\mathbf{k}_1)\tilde{\rho}(\mathbf{k}_2) \tilde{\rho}^*(\mathbf{k}_1 + \mathbf{k}_2) \rangle,
    \label{eq:3ptcorr}
\end{split}
\end{align}
\noindent
where $\mathbf{k}_1$ and $\mathbf{k}_2$ are wave vectors \citep{bernardeau2002large}. The power spectrum in Eq. \ref{eq:2ptcorr} does not capture phase information, while the bispectrum does.

Similar to Eq. \ref{eq:ps} we isotropically sum the bispectrum over modes with fixed scalar wavenumbers ($k_1$, $k_2$, $k_3$), where $k_3 = |\mathbf{k}_1 + \mathbf{k}_2|$. 
Rather than parametrizing by wavenumber $k_3$, we do so by the opening triangle angle $\theta$ between wave vectors with side lengths ($k_1$, $k_2$). In particular, following the conventions in Eq. \ref{eq:ps}, we compute the isotropic bispectrum
\begin{equation}
B(k_{1},k_{2},\theta)= \sum\limits_{\Omega}
\tilde{\rho}(\mathbf{k}_{1}) \tilde{\rho}(\mathbf{k}_{2}) \tilde{\rho}^{*}(\mathbf{k}_1 + \mathbf{k}_2),
\label{eq:bispectra}
\end{equation}
where $\Omega = \Omega(k_1, k_2, \theta)$ is the set of ($\mathbf{k}_1, \mathbf{k}_2$) such that $|\mathbf{k}_1| \in [k_1, k_1+\Delta k)$, $|\mathbf{k}_2| \in [k_2, k_2+\Delta k)$, and $\arccos(\hat{\mathbf{k}}_1\cdot\hat{\mathbf{k}}_2) \in [\theta, \theta+\Delta\theta)$. Equation \ref{eq:bispectra} is a real-valued function since complex conjugate wave vector pairs ($\mathbf{k}_1, \mathbf{k}_2$) and ($-\mathbf{k}_1, -\mathbf{k}_2$) are both contained in $\Omega$. 
It is important to note that the direction of the external magnetic field $\mathbf{B}_0$ introduces a non-trivial dependence in the bispectrum on two angular variables describing triangle orientation with respect to $\mathbf{B}_0$. 
A future topic could be to compute power spectrum and bispectrum multipoles in this dependence \citep{Scoccimarro2015FastEF}.

\begin{figure}
    \center
    \includegraphics[width=\textwidth/2]{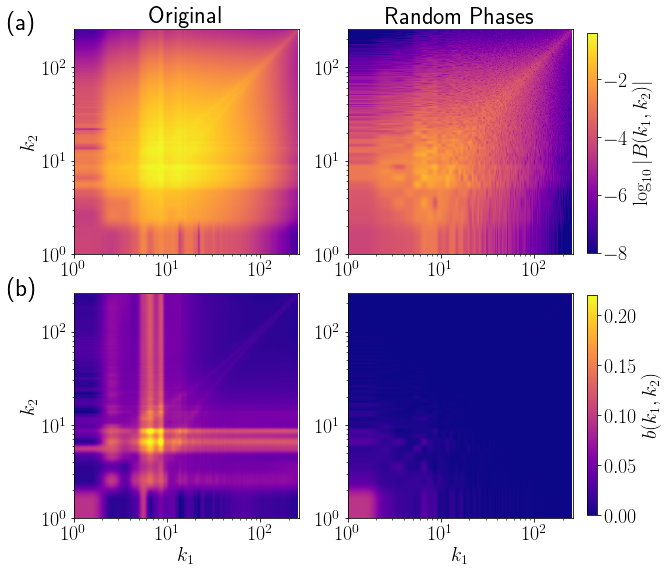}
    \caption{(a) The left depicts the bispectrum for $N_{\rm res} = 512^3$ and $k_{\rm drive} = 7$, while the right depicts the bispectrum of the same data, where the phases of the FFT have been scrambled before calculating. (b) The same setup as (a) for the bicoherence. For the phase-scrambled case, $B$ subtly preserves a signal while $b \approx 0$. This calculation gives intuition for how the bicoherence measures phase coupling and demonstrates the utility of higher-order statistics.}
    \label{fig:Scramble}
\end{figure}

In this work we also introduce the bicoherence index, a normalization of the bispectral amplitudes that measures phase coupling. We define the bicoherence as
\begin{align}
	\begin{split}
		b(k_{1},k_{2},\theta) &= \frac{
		|B(k_1, k_2, \theta)|}
		{\sum\limits_{\Omega}
		|\tilde{\rho}(\mathbf{k}_{1}) \tilde{\rho}(\mathbf{k}_{2}) \tilde{\rho}^*(\mathbf{k}_1 + \mathbf{k}_2)|}.
    \end{split}
\label{eq:bicoherence}
\end{align}
The bicoherence satisfies $0 \leq b \leq 1$. Decomposing $\rho(\mathbf{k}) = A(\mathbf{k})\exp{i\phi(\mathbf{k})}$ with $A(\mathbf{k})$ real, the summand of the bispectrum decomposes into an amplitude term $A(\mathbf{k}_{1}) A(\mathbf{k}_{2}) A(\mathbf{k}_1 + \mathbf{k}_2)$ and a phase term $\phi(\mathbf{k}_1)+ \phi(\mathbf{k}_2)-\phi(\mathbf{k}_1 + \mathbf{k}_2)$. Thus, the bicoherence normalization sums only over the amplitudes. If the phase term is a constant over $\Omega$, then $b = 1$. On the other hand if the phases are random and uniform over $(0, 2\pi]$, then $b \approx 0$. Figure \ref{fig:Scramble} demonstrates how scrambling the phases of an FFT drives the bicoherence to zero. There can be many choices for normalizing bispectra, but here we choose one with a simple geometric interpretation with regard to our spatial averaging.

We reduce dimensionality further by summing over all $\theta$ contributions for fixed $(k_1, k_2)$. 
More specifically, we consider $B = B(k_1, k_2)$ and $b = b(k_1, k_2)$, where we define $\Omega$ to have a fixed $(k_1, k_2)$ but any $\theta$, i.e. $\Omega = \Omega(k_1, k_2)$. Summing out the $\theta$ dependency is sufficient for our purposes here. That said, our code implementation described in the Appendix can quickly calculate an entire distribution over $\theta$, which is important as many implementations of the bispectrum either average over angles or only calculate for a narrow band of angles to reduce the amount of information. Allowing for variable triangle configurations could be important when one also varies the sonic and the Alfv\'enic Mach numbers $\mathcal{M}_s$ and $\mathcal{M}_A$, much like in \cite{Burkhart09a}

Related to the bispectrum is the three-point correlation function (3PCF). The 3PCF is the inverse Fourier transform of the bispectrum, i.e., it is the real-space analogy of the relationship between the two-point correlation function and the power spectrum.  A fast multipole expansion for the 3PCF in opening triangle angle $\theta$ was applied to MHD turbulence simulations in \citet{Portillo2018ApJ...862..119P}, which found a mild sensitivity to the sonic and Alfv\'enic Mach numbers.

The bispectrum $B(k_1, k_2)$ has been applied to simulations and observations \citep{Burkhart09a,Portillo2018ApJ...862..119P} in the context of characterizing non-Gaussian features arising from compressible turbulence. These studies found that the bispectrum is a sensitive diagnostic
for both the sonic and the Alfv\'enic Mach numbers.  More importantly, the bispectrum can describe the behavior of non-linear mode correlation across
spatial scales.

The bispectrum characterizes and searches for
non-linear interactions and non-Gaussianity, which makes it a useful
technique for studies of supersonic super-Alfv\'enic MHD turbulence in the ISM and solar wind due to the fact that as turbulent eddies evolve, they transfer energy from 
large scales to small scales, generating a hierarchical turbulence cascade.
For incompressible Kolmogorov-type flows, this can be expressed as $k_1 \approx k_2 = k$ and $k_3 \approx  2k$.
Non-linear  interactions take place more strongly in 
compressible and magnetized flows, i.e., $k_1 \ne k_2$.
The bispectrum as well as other three-point statistics can characterize these non-linear interactions.

A complex-valued version of the bispectrum has been used to study time-series correlations. Its phase content has been shown to contain rich information about wave shape \citep{BiphaseExplained} and direction of energy transfer in a fluid \citep{BiphaseEnergyTransfer}. Other recent works define a complex-valued bispectrum over wave vector and frequency space \citep{schmidt2020bispectral}.

In the following we simply refer to $P(k)$ as the power spectrum and $B(k_1, k_2)$ as the bispectrum. In the appendices we describe the specifics of the implementation, which allows us to calculate density bispectra for a 3D simulation using $N_{\rm res} = 1024^3$ in an hour and for a 2D column density map $N_{\rm res} = 1024^2$ in under a minute (in double-precision on a Nvidia A100 40GB GPU). In Appendix \ref{appendix:estimator}, we describe a Monte Carlo estimator for calculating Eqs. \ref{eq:bispectra} and \ref{eq:bicoherence}. We demonstrate numerical convergence in Appendix \ref{appendix:convergence} and describe the open-source implementation in Appendix \ref{appendix:implementation}. Finally, we describe a caveat to summing over $\theta$ in Appendix \ref{appendix:corrections} and include a discussion of the advantages and disadvantages of our method compared to recent algorithmic developments in Appendix \ref{appendix:fftestimator}.

\subsection{Averaging the Bispectrum}

The bispectrum $B(k_1, k_2)$ is more difficult to interpret than the power spectrum $P(k)$ as it is represented by a 2D map of wavenumbers. To address this issue, \citet{Burkhart2016ApJ...827...26B} introduced averaging over bispectral contours to reduce the information
provided by the bispectral amplitudes.

To compactly represent our data, we define the sum over lines of constant $k_1 = k$

\begin{align}
	\langle B(k_1, k_2) \rangle_k &= \sum_{k_1} B(k_1, k) \\
    \langle b(k_1, k_2) \rangle_k &= \sum_{k_1} b(k_1, k).
\label{eq:avg}
\end{align}

Choosing to sum over constant $k_1$ rather than $k_2$ is arbitrary because $B(k_1, k_2) = B(k_2, k_1)$. This particular way to reduce the information contained in the bispectrum and bicoherence conveniently highlights the signals that we see in our data. It is also a natural way to average while capturing correlations for a fixed $k$. 
Geometrically, it fixes a triangle side length $k_1 = k$ and sums together all triangles with varying side length $k_2$.

\section{Studying Driving Scales}
\label{sec:res}

In this paper we calculate the bispectrum for the density field $\rho(\mathbf{x}, t)$ of our simulations, where the dynamics has reached a statistical steady state. Specifically, we compute the bispectrum of $\tilde{\rho}(\mathbf{k}) =  \mathcal{F}[\rho(\mathbf{x}) - \bar{\rho}]$, where $\bar{\rho}$ is the mean of the density field and $\mathcal{F}$ denotes the Fourier transform. Subtracting by the mean has the effect of setting $\tilde{\rho}(\mathbf{k} = 0) = 0$, which highlights off-diagonal components of the bispectrum in the contour plots of Figure \ref{fig:MainFigure}, for example.

Figure \ref{fig:MainFigure}(a) illustrates an example density field for $k_{\rm{\rm drive}} = 5$ and $N_{\rm res} = 1024^3$. Its bicoherence $b$ in Fig. \ref{fig:MainFigure}(b) is strongest for fixed $k_1 = k_{\rm drive}$ for $k_2 \geq k_{\rm drive}$. In other words, scales in the turbulent cascade are phase-coupled to the driving scale.

In the following sections we consider the results of our $N_{\rm res} = 1024^3$ simulations for varying $k_{\rm drive}$.

\subsection{Density}
\label{sec:dens}

\begin{figure}
    \center
    \includegraphics[width=\textwidth/2]{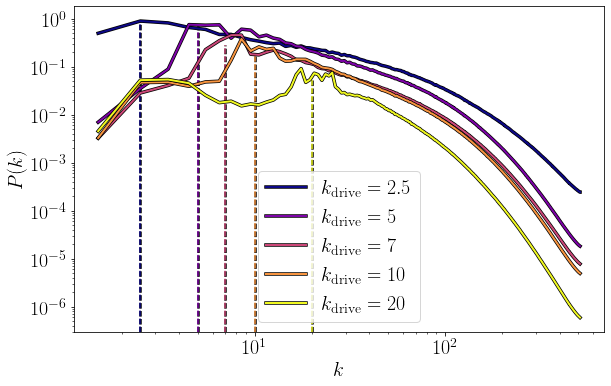}
    \caption{The power spectra $P(k)$ of density for varying driving scale $k_{\rm drive}$ with resolution $N_{\rm res} = 1024^3$. The spectra are peaked near the driving scales. Amplitudes are also enhanced for small $k$, which is analyzed further in Fig. \ref{fig:MhdVsHydro}.}
    \label{fig:Pspec_Dens}
\end{figure}

Figure \ref{fig:Pspec_Dens} shows the density power spectra for a range of driving scales. These alone, using only amplitude data, display evidence of the driving scales.

\begin{figure*}
    \center
    \includegraphics[width=\textwidth]{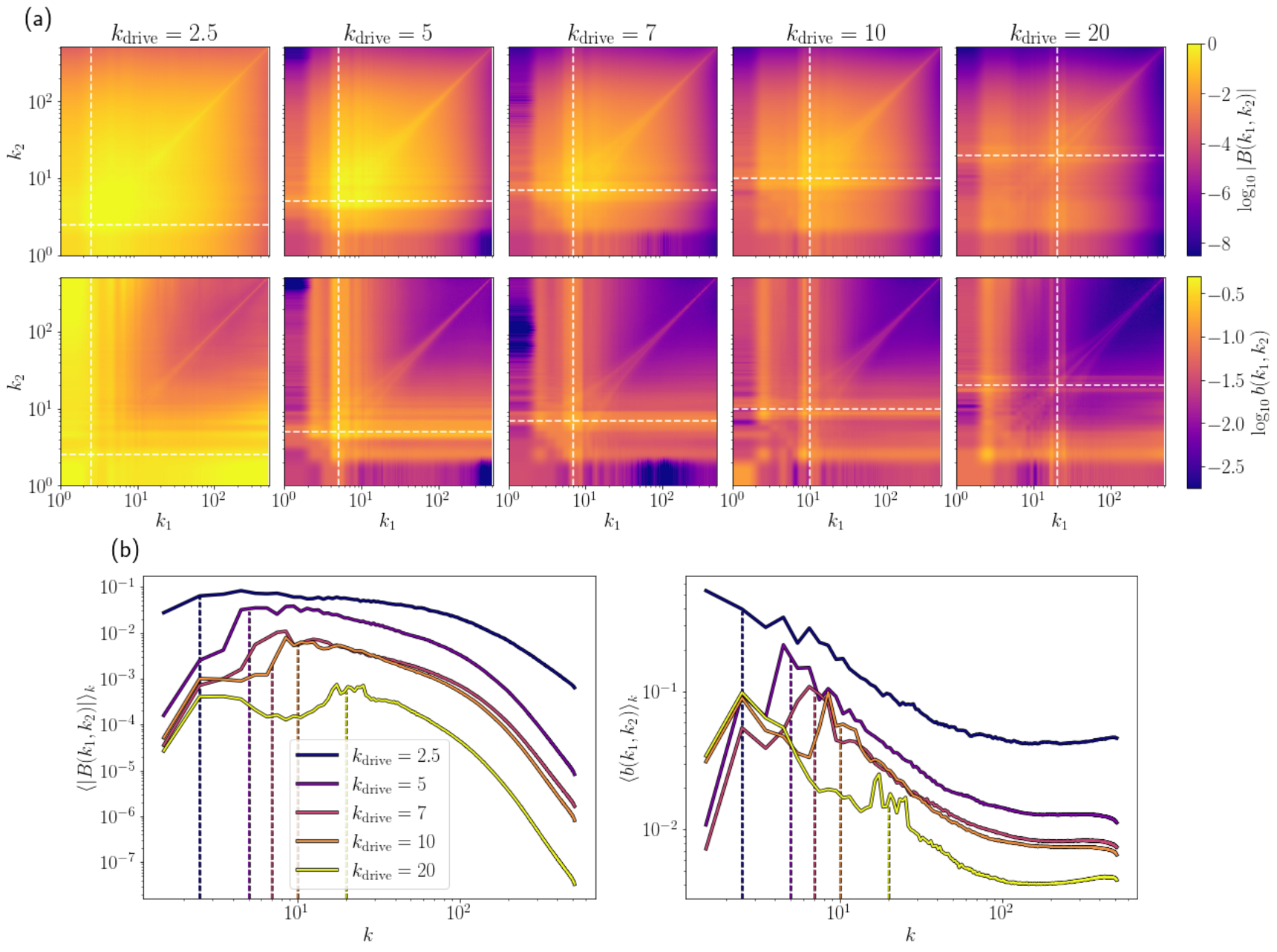}
    \caption{(a) Bispectral ampltudes $|B(k_1, k_2)|$ and bicoherence $b(k_1, k_2)$ plotted up to the Nyquist frequency for varying driving scale at $N_{\rm res} = 1024^3$. Similar to Fig \ref{fig:MainFigure}, there is the L-shaped signature of the turbulence driving scale--however of decreasing phase coupling strength as indicated by $b$. For $k_{\rm drive} = 7, \ 10, \ 20$ there are additional correlations in a very similar fashion not present in classical turbulence (see Fig. \ref{fig:MhdVsHydro}). (b) Averaged bispectral amplitudes and bicoherence (see Eq. \ref{eq:avg}) of (a). The bicoherence more strongly picks out the driving scale by highlighting Fourier phase correlations.}
    \label{fig:BispecVsBicoh_Dens}
\end{figure*}

In Fig. \ref{fig:BispecVsBicoh_Dens}(a), the bicoherence and the bispectrum as defined in Eq. \ref{eq:avg} can also reliably pick out driving scales by illustrating the scales present in the turbulence cascade.
However, as also displayed by Fig. \ref{fig:MainFigure}(b) the bicoherence shows a stronger driving scale signature than the bispectrum.
The larger $k_{\rm drive}$ cases also show additional correlations at long-waves in the power spectra of Fig. \ref{fig:Pspec_Dens} and averaged bispectra and bicoherence of Fig. \ref{fig:BispecVsBicoh_Dens}(b).

\begin{figure}
    \center
    \includegraphics[width=\textwidth/2]{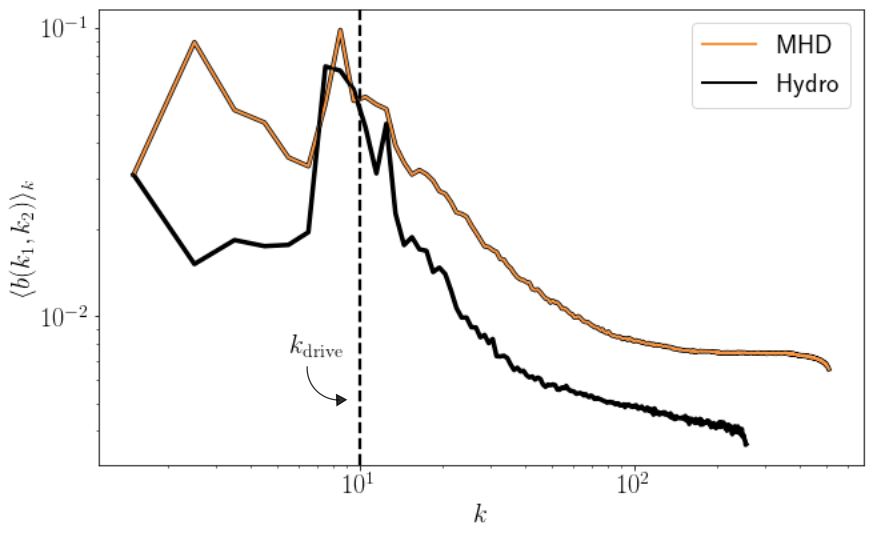}
    \caption{Comparison of the averaged bicoherence for MHD turbulence $\mathcal{M}_A = 0.7$ and purely hydrodynamic turbulence. Both cases detect $k_{\rm drive}$, but the MHD case also detects correlations from the  magnetic field.}
    \label{fig:MhdVsHydro}
\end{figure}

Figure \ref{fig:MhdVsHydro} suggests a reason for these additional correlations. Here, we compare the averaged bicoherence for MHD turbulence and hydrodynamic turbulence at $k_{\rm drive} = 10$.
The hydrodynamic case does not show correlations, hinting that the external magnetic field in MHD turbulence induces mode-mode coupling. We discuss the significance of these magnetic field correlations in Section \ref{sec:dis}.

\subsection{Column density}
\label{sec:cdens}

In this section we show results for integrated column density along axes perpendicular and parallel to the background magnetic field $\mathbf{B}_0$. Results from column density can have applications for observational studies, particularly here for a low-gravity limit. An important future study would study driving scales for models incorporating both turbulence and gravity.

\begin{figure}
    \center
    \includegraphics[width=\textwidth/2]{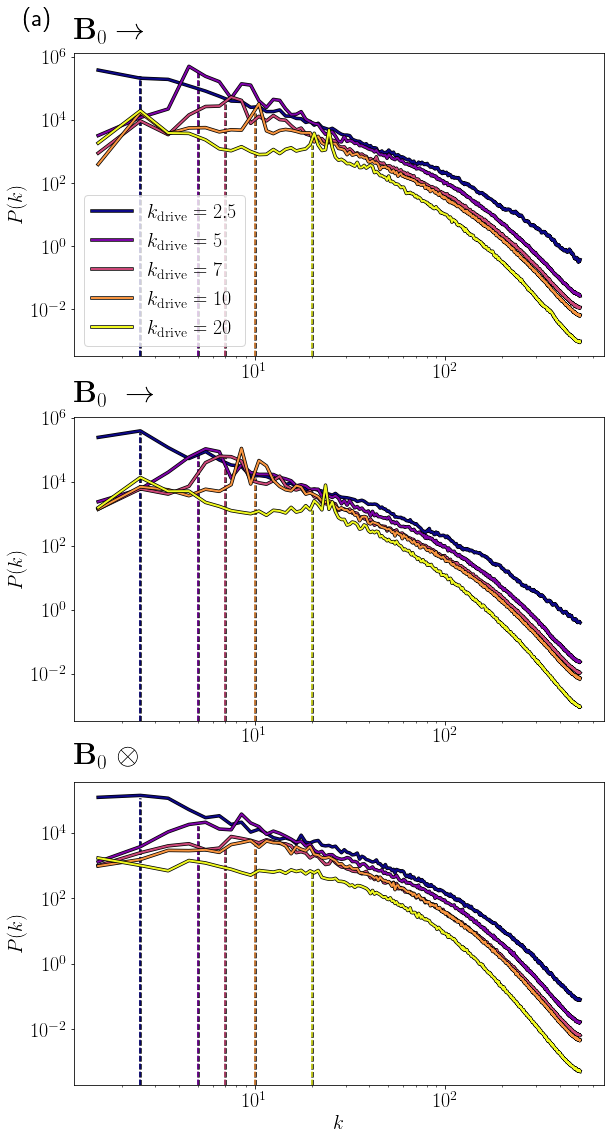}
    \caption{(a) The power spectra $P(k)$ of column density projected along the $z$ direction, perpendicular to the background magnetic field. (b) $P(k)$ along the $y$ direction. Perpendicular to the magnetic field, the power spectrum can capture the driving scale similar to Fig \ref{fig:Pspec_Dens}. Along these two axes, there are also visible long-wave enhancements as described in Fig. \ref{fig:MhdVsHydro}. (c) The projection along the background magnetic field. Driving scale signatures and long-wave enhancements disappear along this axis. For consistency, we have labeled the direction of $\mathbf{B}_0$ in (a), (b), and (c) the same as the column densities in Fig. \ref{fig:CDensPerp} and Fig. \ref{fig:CDensAlong}.}
    \label{fig:Pspec_CDens}
\end{figure}

\begin{figure*}
    \center
    \includegraphics[width=\textwidth]{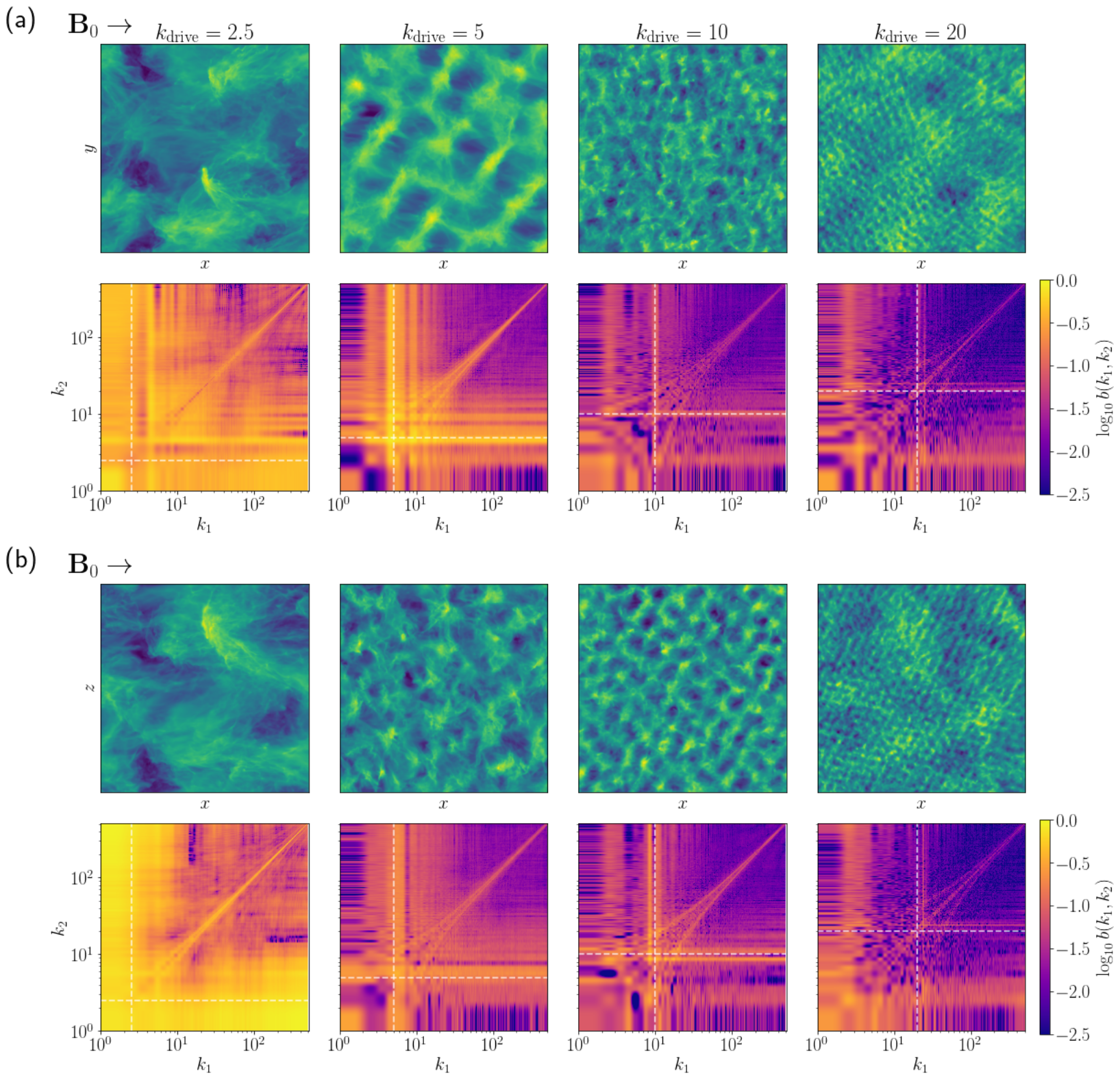}
    \caption{
    Column density projected along axes perpendicular to the background magnetic field and the associated bicoherence. (a) depicts the projection along the $z$ axis and (b) the $y$ axis.
    The driving scales are identifiable just as in Fig. \ref{fig:BispecVsBicoh_Dens}(a). The $k_{\rm drive} = 10$ and $20$ cases also clearly depict the long-wave magnetic field correlations along with those from turbulence driving.}
    \label{fig:CDensPerp}
\end{figure*}

\begin{figure*}
    \center
    \includegraphics[width=\textwidth]{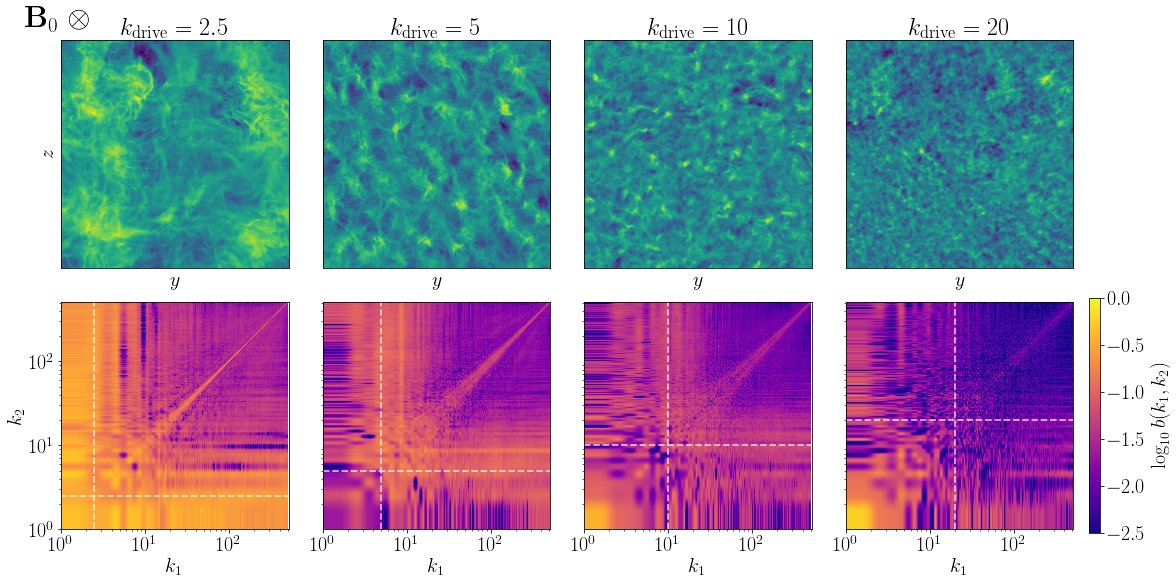}
    \caption{
    Column density and its associated bicoherence along the $x$ direction, the axis parallel to the background magnetic field. The correlation enhancement in the bicoherence are much less apparent than Fig. \ref{fig:CDensPerp} and is particularly hard to see for small-scale driving (i.e. $k_{\rm drive} = 10$ and $20$). The long-wave correlations due to the magnetic field do not appear at all. Fig \ref{fig:Pspec_CDens}(c) demonstrates how the power spectrum virtually has no driving scale signal, which is not the case here.}
    \label{fig:CDensAlong}
\end{figure*}

Column density projections perpendicular and parallel to the background magnetic field $\mathbf{B}_0$ are expected to have different statistics. In MHD turbulence, projections perpendicular to $\mathbf{B}_0$ are expected to look anisotropic and parallel projections isotropic.

Figure \ref{fig:Pspec_CDens} shows the power spectrum across driving scales of column density.
In Fig. \ref{fig:Pspec_CDens}(a), looking perpendicular to $\mathbf{B}_0$ along the $z$ axis, there is an indication of $k_{\rm drive}$ for all cases. However, it is particularly unclear for long-wave driving, $k_{\rm drive} = 2.5, \ 5$. The story is similar along the $y$ axis in \ref{fig:Pspec_CDens}(b). The signal is fainter than for the power spectra of the whole 3D field in Fig. \ref{fig:Pspec_Dens}.
Figure \ref{fig:Pspec_CDens}(c) is density projected parallel to $\mathbf{B}_0$. Here, there is less indication of driving scale for all $k_{\rm drive}$. 

Studying the bispectrum and bicoherence is more reliable in extracting driving scales. Figures \ref{fig:CDensPerp}(a) and (b) show column density images along both directions perpendicular to $\mathbf{B}_0$ for $k_{\rm drive} = 2.5, \ 5, \ 10,$ and $20$. Each image has its bicoherence illustrated beneath.

Figures \ref{fig:CDensPerp}(a) and \ref{fig:CDensPerp}(b) show driving scale signatures more reliable than that of the power spectrum in Fig. \ref{fig:Pspec_CDens}(a) and \ref{fig:Pspec_CDens}(b).
For example, $k_{\rm drive} = 5$ is especially clear for the bicoherence and less so for the power spectrum. 
The amplitude enhancement in the power spectrum could be easily washed out by smoothing in observations too.

Notably, the nature of the signal between the bicoherence and power spectrum is very different. 
The stronger power spectral amplitudes are a result of momentum injection in a very narrow band of Fourier modes.
Realistic driving is less simple and requires more robust metrics beyond simple Fourier amplitude enhancement.
The bicoherence is able to recognize self-similarity across scales, a physical phenomenon of turbulence.

Figure \ref{fig:CDensAlong} illustrates for column density integrated along $\mathbf{B}_0$. 
Like in Fig. \ref{fig:Pspec_CDens}(c), it is difficult to pick out the driving scale.
The column density maps (particularly for $k_{\rm drive} = 10$ and $20$) exhibit more line-of-sight confusion than those in Fig. \ref{fig:CDensPerp}. The smaller scale features do not clearly show a precise preferred scale.

However, the bispectrum is a more robust statistic for analyzing driving scales even along this axis.
Correlation enhancements (particularly for $k_{\rm drive} = 2.5$ and $5$) appear to have higher signal-to-noise than the power spectrum.

Interestingly, the long-wave correlations (explained in Fig. \ref{fig:MhdVsHydro}) disappear when looking along the magnetic field. This is visible in both the power spectrum in Fig. \ref{fig:Pspec_CDens}(c) and the bicoherence in Fig. \ref{fig:CDensAlong}.

\section{Discussion}
\label{sec:dis}

In this paper we have demonstrated the utility of the bispectrum (and related bicoherence) for determining driving scales of turbulence in the interstellar medium.  The driving scale is an essential physical aspect of a turbulent cascade, where energy can begin to cascade either to small (for a forward cascade) or to large scales (in an inverse cascade). Our study paves the way for future applications of the bispectrum and bicoherence for observational data.  Future investigation of the effects of radiative transfer to study particular line tracers (e.g. CO, HI) as well as dust extinction/emission will be performed to further assess the application of our technique for observational data.  We also suggest the study of multiple driving scales and more physical driving mechanisms, such as supernova driven turbulence and outflow driven turbulence. We are conducting a follow up study to address these.

The bispectrum includes phase information, whereas the traditional Fourier power spectrum only contains the amplitudes. The addition of phases provides more structural information that amplitude alone does not capture. The power spectrum more reliably provide the driving scale when dealing with 3D quantities such as density and kinetic energy, but when applied to projected quantities, such as column density, it is no longer able to distinguish the driving scale over noise and effects of the changing magnetic orientation.  Interestingly, our results show that the phase information is primarily tracing the driving scale, since the bicoherence is more sensitive than the bispectrum in terms of picking out the driving scale.  This is because the bispectrum includes both phase and amplitude information while the bicoherence normalizes out the Fourier phases.

Furthermore, quantifying the spatial structure of the ISM with statistical diagnostics that are sensitive the Fourier phases allows the full information contained in images produced by numerical simulations to be more directly compared with observations. 
The importance of phase information has become clear recently with the rising interest and use of machine learning algorithms for classifying turbulence. For example, in \cite{Peek2019} a 
 trained convolutional neural network was able to tell sub-Alfv\'enic vs.
super-Alfv\'enic simulations apart with about 98\% accuracy due to phase information alone.
The network was not presented with any of the power spectra information and the only
information available was embedded in the Fourier phase domain. This experiment has shown that there is a large amount
of valuable information in Fourier phase space which is not available
to the power spectrum.
Phase information is also of general interest in the MHD turbulence community. Non-linear interactions among MHD waves (i.e., slow, fast, and Alfv\'en) produce finite
correlation of the wave phases. As a result, several other statistical studies have suggested using phase information present in MHD turbulence to understand non-linear mode interactions and plasma turbulence \cite{Burkhart2016ApJ...827...26B,Portillo2018ApJ...862..119P}.

We also found that averaging the bispectrum and bicoherence is a very useful post-processing step that allows the bispectral wave vector parameter space to be distilled down to a simpler representation.  Our Fig. \ref{fig:BispecVsBicoh_Dens} reveals this averaging demonstrates not only the scale of the driving but also a sensitivity to the presence of a magnetic field.  A long-wave mean magnetic field produces correlations that dominate the small wavenumbers, whereas our hydrodynamic simulations do not exhibit this enhancement of correlation. This is likely due to non-local effects present in the MHD turbulence cascade. Non-locality in MHD turbulence is much more pronounced than that of hydrodynamic turbulence \citep{2010ApJ...725.1786C,2010ApJ...722L.110B}.
In a forward-cascade, hydrodynamic (HD) turbulence, energy moves from large to small scales as kinetic energy  is transferred to smaller eddies by shearing motions of other eddies.
Locality in this framework means that interactions between eddies of similar size dominate such that, in Fourier space, a Fourier mode at a wavenumber $k$ interacts primarily with other modes having similar wavenumbers and energy is transferred to larger wavenumbers. 
MHD turbulence has non-local effects in the form of large-scale shearing motion  (which does not promote energy transfer) and non-local energy transfer.  This can explain why we see additional mode correlation at scales larger than the driving scales. Further study of the bispectrum of kinetic energy is warranted in order to confirm the effect we see is due to non-locality effects. 
The effect of the magnetic field on the bicoherence and phase coupling should be studied further as a parameter based investigation where the strength of the magnetic field is altered.

\section{Conclusions}
\label{sec:con}

In this work we present an adaptable and easy-to-use Monte Carlo GPU/CPU bispectrum and bicoherence implementation. Our code is made publicly available for general use. We describe the implementation and link to the codebase in Appendix \ref{appendix:implementation}.
We have applied the bispectrum and bicoherence to isothermal MHD driven turbulence simulations which have different driving scales from $k = 2.5$ to $k = 20$.

We find that:
\begin{itemize}
    \item The power spectrum, bicoherence, and bispectrum are all sensitive to the driving scale of compressible MHD turbulence when applied to the density field.
    \item The bicoherence of MHD turbulence density fields reveal that Fourier phase correlations are responsible for outlining the scales present in the turbulent cascade.
    \item When studying projected density (Column density)  the bicoherence retains  signatures of the driving scales. This opens up  applications of the bicoherence to astrophysical observations in order to determine the driving scale of turbulence.
    \item The presence of an external magnetic field induces additional mode-mode coupling in the MHD cascade that are not present in a hydrodynamic cascade. The enhanced mode coupling is clear at large scales and can be larger than the driving scale of the turbulence. This may be a signature of Alfv\'en waves traveling along field lines and  inducing an additional cascade at larger scales.
\end{itemize}

\section*{acknowledgments}

Authors are grateful for useful comments and discussion with Oliver Philcox. The analysis for this work was performed on the Flatiron Institute Rusty cluster. 
B.B. is grateful for funding support by the Simons Foundation, Sloan Foundation, and the Packard Foundation.

\bibliography{refsMASTER.bib}

\begin{thebibliography}{67}
\expandafter\ifx\csname natexlab\endcsname\relax\def\natexlab#1{#1}\fi

\bibitem[{Banerjee {et~al.}(2007)Banerjee, Klessen, \& Fendt}]{Banerjee_2007}
Banerjee, R., Klessen, R.~S., \& Fendt, C. 2007, The Astrophysical Journal,
  668, 1028

\bibitem[{{Beresnyak} \& {Lazarian}(2010)}]{2010ApJ...722L.110B}
{Beresnyak}, A. \& {Lazarian}, A. 2010, \apjl, 722, L110

\bibitem[{Bernardeau {et~al.}(2002)Bernardeau, Colombi, Gaztanaga, \&
  Scoccimarro}]{bernardeau2002large}
Bernardeau, F., Colombi, S., Gaztanaga, E., \& Scoccimarro, R. 2002, Physics
  reports, 367, 1

\bibitem[{{Bialy} \& {Burkhart}(2020)}]{bialy2020}
{Bialy}, S. \& {Burkhart}, B. 2020, \apjl, 894, L2

\bibitem[{{Bialy} {et~al.}(2017){Bialy}, {Burkhart}, \&
  {Sternberg}}]{bialy2017ApJ...843...92B}
{Bialy}, S., {Burkhart}, B., \& {Sternberg}, A. 2017, \apj, 843, 92

\bibitem[{{Burkhart} {et~al.}(2009){Burkhart}, {Falceta-Gon{\c c}alves},
  {Kowal}, \& {Lazarian}}]{Burkhart09a}
{Burkhart}, B., {Falceta-Gon{\c c}alves}, D., {Kowal}, G., \& {Lazarian}, A.
  2009, \apj, 693, 250

\bibitem[{{Burkhart} \& {Lazarian}(2016)}]{Burkhart2016ApJ...827...26B}
{Burkhart}, B. \& {Lazarian}, A. 2016, \apj, 827, 26

\bibitem[{{Burkhart} {et~al.}(2013){Burkhart}, {Lazarian}, {Goodman}, \&
  {Rosolowsky}}]{burkhart13}
{Burkhart}, B., {Lazarian}, A., {Goodman}, A., \& {Rosolowsky}, E. 2013, \apj,
  770, 141

\bibitem[{{Burkhart} {et~al.}(2015){Burkhart}, {Lee}, {Murray}, \&
  {Stanimirovi{\'c}}}]{Burkhart2015ApJ...811L..28B}
{Burkhart}, B., {Lee}, M.-Y., {Murray}, C.~E., \& {Stanimirovi{\'c}}, S. 2015,
  \apjl, 811, L28

\bibitem[{{Burkhart} \& {Mocz}(2019)}]{BurkhartMocz2019}
{Burkhart}, B. \& {Mocz}, P. 2019, \apj, 879, 129

\bibitem[{{Burkhart} {et~al.}(2017){Burkhart}, {Stalpes}, \&
  {Collins}}]{Burkhart2017ApJ...834L...1B}
{Burkhart}, B., {Stalpes}, K., \& {Collins}, D.~C. 2017, \apjl, 834, L1

\bibitem[{{Chepurnov} {et~al.}(2015){Chepurnov}, {Burkhart}, {Lazarian}, \&
  {Stanimirovic}}]{chepurnov15}
{Chepurnov}, A., {Burkhart}, B., {Lazarian}, A., \& {Stanimirovic}, S. 2015,
  \apj, 810, 33

\bibitem[{{Cho}(2010)}]{2010ApJ...725.1786C}
{Cho}, J. 2010, \apj, 725, 1786

\bibitem[{{Collins} {et~al.}(2012){Collins}, {Kritsuk}, {Padoan}, {Li}, {Xu},
  {Ustyugov}, \& {Norman}}]{Collins12}
{Collins}, D.~C., {Kritsuk}, A.~G., {Padoan}, P., {Li}, H., {Xu}, H.,
  {Ustyugov}, S.~D., \& {Norman}, M.~L. 2012, \apj, 750, 13

\bibitem[{Cui \& Jacobi(2021)}]{BiphaseEnergyTransfer}
Cui, G. \& Jacobi, I. 2021, Phys. Rev. Fluids, 6, 014604

\bibitem[{Cunningham {et~al.}(2006)Cunningham, Frank, \&
  Blackman}]{Cunningham_2006}
Cunningham, A.~J., Frank, A., \& Blackman, E.~G. 2006, The Astrophysical
  Journal, 646, 1059–1069

\bibitem[{{Cunningham} {et~al.}(2009){Cunningham}, {Frank}, {Carroll},
  {Blackman}, \& {Quillen}}]{2009ApJ...692..816C}
{Cunningham}, A.~J., {Frank}, A., {Carroll}, J., {Blackman}, E.~G., \&
  {Quillen}, A.~C. 2009, \apj, 692, 816

\bibitem[{{Dziourkevitch} {et~al.}(2004){Dziourkevitch}, {Elstner}, \&
  {R{\"u}diger}}]{2004A&A...423L..29D}
{Dziourkevitch}, N., {Elstner}, D., \& {R{\"u}diger}, G. 2004, \aap, 423, L29

\bibitem[{{Eisenstein} {et~al.}(2007){Eisenstein}, {Seo}, {Sirko}, \&
  {Spergel}}]{2007ApJ...664..675E}
{Eisenstein}, D.~J., {Seo}, H.-J., {Sirko}, E., \& {Spergel}, D.~N. 2007, \apj,
  664, 675

\bibitem[{{Elmegreen} \& {Scalo}(2004)}]{ElmegreenScalo}
{Elmegreen}, B.~G. \& {Scalo}, J. 2004, \araa, 42, 211

\bibitem[{{Federrath}(2015)}]{Federrath2015a}
{Federrath}, C. 2015, \mnras, 450, 4035

\bibitem[{{Forbes} {et~al.}(2014){Forbes}, {Krumholz}, {Burkert}, \&
  {Dekel}}]{Forbes14a}
{Forbes}, J.~C., {Krumholz}, M.~R., {Burkert}, A., \& {Dekel}, A. 2014, \mnras,
  438, 1552

\bibitem[{{Gallegos-Garcia} {et~al.}(2020){Gallegos-Garcia}, {Burkhart},
  {Rosen}, {Naiman}, \& {Ramirez-Ruiz}}]{Gallegos-Garcia:2020:ApJL}
{Gallegos-Garcia}, M., {Burkhart}, B., {Rosen}, A.~L., {Naiman}, J.~P., \&
  {Ramirez-Ruiz}, E. 2020, \apjl, 899, L30

\bibitem[{{Goldbaum} {et~al.}(2015){Goldbaum}, {Krumholz}, \&
  {Forbes}}]{Goldbaum15a}
{Goldbaum}, N.~J., {Krumholz}, M.~R., \& {Forbes}, J.~C. 2015, \apj, 814, 131

\bibitem[{{Goldreich} \& {Sridhar}(1995)}]{GS95}
{Goldreich}, P. \& {Sridhar}, S. 1995, \apj, 438, 763

\bibitem[{{Gressel} {et~al.}(2011){Gressel}, {Elstner}, \&
  {R{\"u}diger}}]{2011IAUS..274..348G}
{Gressel}, O., {Elstner}, D., \& {R{\"u}diger}, G. 2011, in Advances in Plasma
  Astrophysics, ed. A.~{Bonanno}, E.~{de Gouveia Dal Pino}, \& A.~G.
  {Kosovichev}, Vol. 274, 348--354

\bibitem[{{Haverkorn} {et~al.}(2008){Haverkorn}, {Brown}, {Gaensler}, \&
  {McClure-Griffiths}}]{Haverkorn2008}
{Haverkorn}, M., {Brown}, J.~C., {Gaensler}, B.~M., \& {McClure-Griffiths},
  N.~M. 2008, \apj, 680, 362

\bibitem[{{Herron} {et~al.}(2016){Herron}, {Burkhart}, {Lazarian}, {Gaensler},
  \& {McClure-Griffiths}}]{Herron2016ApJ...822...13H}
{Herron}, C.~A., {Burkhart}, B., {Lazarian}, A., {Gaensler}, B.~M., \&
  {McClure-Griffiths}, N.~M. 2016, \apj, 822, 13

\bibitem[{{Hill} {et~al.}(2008){Hill}, {Benjamin}, {Kowal}, {Reynolds},
  {Haffner}, \& {Lazarian}}]{Hill2008}
{Hill}, A.~S., {Benjamin}, R.~A., {Kowal}, G., {Reynolds}, R.~J., {Haffner},
  L.~M., \& {Lazarian}, A. 2008, \apj, 686, 363

\bibitem[{{Kowal} \& {Lazarian}(2007)}]{Kowal07}
{Kowal}, G. \& {Lazarian}, A. 2007, \apjl, 666, L69

\bibitem[{{Kritsuk} {et~al.}(2007){Kritsuk}, {Norman}, {Padoan}, \&
  {Wagner}}]{Kritsuk07a}
{Kritsuk}, A.~G., {Norman}, M.~L., {Padoan}, P., \& {Wagner}, R. 2007, \apj,
  665, 416

\bibitem[{{Krumholz}(2014)}]{krumreview2014}
{Krumholz}, M.~R. 2014, \physrep, 539, 49

\bibitem[{{Krumholz} \& {Burkhart}(2016)}]{2016MNRAS.458.1671K}
{Krumholz}, M.~R. \& {Burkhart}, B. 2016, \mnras, 458, 1671

\bibitem[{{Krumholz} {et~al.}(2018){Krumholz}, {Burkhart}, {Forbes}, \&
  {Crocker}}]{krumholz2018}
{Krumholz}, M.~R., {Burkhart}, B., {Forbes}, J.~C., \& {Crocker}, R.~M. 2018,
  \mnras, 477, 2716

\bibitem[{{Larson}(1981)}]{Lars81}
{Larson}, R.~B. 1981, \mnras, 194, 809

\bibitem[{{Lazarian}(2007)}]{Lazarian07rev}
{Lazarian}, A. 2007, Journal of Quantitative Spectroscopy and Radiative
  Transfer, 106, 225

\bibitem[{{Lazarian}(2009)}]{lazarian09rev}
---. 2009, \ssr, 143, 357

\bibitem[{{Lazarian} \& {Pogosyan}(2006)}]{lp06}
{Lazarian}, A. \& {Pogosyan}, D. 2006, \apj, 652, 1348

\bibitem[{{Lazarian} \& {Vishniac}(1999)}]{LV99}
{Lazarian}, A. \& {Vishniac}, E.~T. 1999, \apj, 517, 700

\bibitem[{{Lazarian} \& {Yan}(2014)}]{LY14}
{Lazarian}, A. \& {Yan}, H. 2014, \apj, 784, 38

\bibitem[{{Le{\~a}o} {et~al.}(2009){Le{\~a}o}, {de Gouveia Dal Pino},
  {Falceta-Gon{\c{c}}alves}, {Melioli}, \& {Geraissate}}]{2009MNRAS.394..157L}
{Le{\~a}o}, M.~R.~M., {de Gouveia Dal Pino}, E.~M., {Falceta-Gon{\c{c}}alves},
  D., {Melioli}, C., \& {Geraissate}, F.~G. 2009, \mnras, 394, 157

\bibitem[{{Mac Low} \& {Klessen}(2004)}]{maclow04}
{Mac Low}, M.-M. \& {Klessen}, R.~S. 2004, Reviews of Modern Physics, 76, 125

\bibitem[{Maccarone(2013)}]{BiphaseExplained}
Maccarone, T.~J. 2013, Monthly Notices of the Royal Astronomical Society, 435,
  3547

\bibitem[{Martizzi {et~al.}(2015)Martizzi, Faucher-Giguère, \&
  Quataert}]{10.1093/mnras/stv562}
Martizzi, D., Faucher-Giguère, C.-A., \& Quataert, E. 2015, Monthly Notices of
  the Royal Astronomical Society, 450, 504

\bibitem[{{McKee} {et~al.}(2010){McKee}, {Li}, \& {Klein}}]{McKee10b}
{McKee}, C.~F., {Li}, P.~S., \& {Klein}, R.~I. 2010, \apj, 720, 1612

\bibitem[{{McKee} \& {Ostriker}(2007)}]{Mckee_Ostriker2007}
{McKee}, C.~F. \& {Ostriker}, E.~C. 2007, \araa, 45, 565

\bibitem[{{Mendel}(1991)}]{75086}
{Mendel}, J.~M. 1991, Proceedings of the IEEE, 79, 278

\bibitem[{{Offner} \& {Arce}(2015)}]{2015ApJ...811..146O}
{Offner}, S. S.~R. \& {Arce}, H.~G. 2015, \apj, 811, 146

\bibitem[{{Padoan} {et~al.}(2017){Padoan}, {Haugb{\o}lle}, {Nordlund}, \&
  {Frimann}}]{padoan2017ApJ...840...48P}
{Padoan}, P., {Haugb{\o}lle}, T., {Nordlund}, {\AA}., \& {Frimann}, S. 2017,
  \apj, 840, 48

\bibitem[{{Padoan} {et~al.}(1997){Padoan}, {Nordlund}, \& {Jones}}]{padoan97}
{Padoan}, P., {Nordlund}, A., \& {Jones}, B.~J.~T. 1997, \mnras, 288, 145

\bibitem[{{Padoan} {et~al.}(2016){Padoan}, {Pan}, {Haugb{\o}lle}, \&
  {Nordlund}}]{2016ApJ...822...11P}
{Padoan}, P., {Pan}, L., {Haugb{\o}lle}, T., \& {Nordlund}, {\r{A}}. 2016,
  \apj, 822, 11

\bibitem[{Pearson \& Samushia(2018)}]{Pearson_2018}
Pearson, D.~W. \& Samushia, L. 2018, Monthly Notices of the Royal Astronomical
  Society, 478, 4500–4512

\bibitem[{{Peek} \& {Burkhart}(2019)}]{Peek2019}
{Peek}, J.~E.~G. \& {Burkhart}, B. 2019, \apjl, 882, L12

\bibitem[{{Pingel} {et~al.}(2018){Pingel}, {Lee}, {Burkhart}, \&
  {Stanimirovi{\'c}}}]{2018ApJ...856..136P}
{Pingel}, N.~M., {Lee}, M.-Y., {Burkhart}, B., \& {Stanimirovi{\'c}}, S. 2018,
  \apj, 856, 136

\bibitem[{{Portillo} {et~al.}(2018){Portillo}, {Slepian}, {Burkhart},
  {Kahraman}, \& {Finkbeiner}}]{Portillo2018ApJ...862..119P}
{Portillo}, S. K.~N., {Slepian}, Z., {Burkhart}, B., {Kahraman}, S., \&
  {Finkbeiner}, D.~P. 2018, \apj, 862, 119

\bibitem[{Regan(2017)}]{Regan2017Estimators}
Regan, D. 2017, Journal of Cosmology and Astroparticle Physics, 2017

\bibitem[{{Santos-Lima} {et~al.}(2010){Santos-Lima}, {Lazarian}, {de Gouveia
  Dal Pino}, \& {Cho}}]{Sant10}
{Santos-Lima}, R., {Lazarian}, A., {de Gouveia Dal Pino}, E.~M., \& {Cho}, J.
  2010, \apj, 714, 442

\bibitem[{{Saydjari} {et~al.}(2021){Saydjari}, {Portillo}, {Slepian},
  {Kahraman}, {Burkhart}, \& {Finkbeiner}}]{2021ApJ...910..122S}
{Saydjari}, A.~K., {Portillo}, S. K.~N., {Slepian}, Z., {Kahraman}, S.,
  {Burkhart}, B., \& {Finkbeiner}, D.~P. 2021, \apj, 910, 122

\bibitem[{{Schlickeiser}(2002)}]{Schlickeiser02}
{Schlickeiser}, R. 2002, {Cosmic Ray Astrophysics}, ed. R.~{Schlickeiser}

\bibitem[{Schmidt(2020)}]{schmidt2020bispectral}
Schmidt, O.~T. 2020, Nonlinear Dynamics, 102, 2479

\bibitem[{Scoccimarro(2015)}]{Scoccimarro2015FastEF}
Scoccimarro, R. 2015, Physical Review D, 92, 083532

\bibitem[{{Slepian} \& {Eisenstein}(2015)}]{Slepian15_alg}
{Slepian}, Z. \& {Eisenstein}, D.~J. 2015, \mnras, 454, 4142

\bibitem[{{Slepian} \& {Eisenstein}(2016)}]{Slepian16_alg_WFTs}
---. 2016, \mnras, 455, L31

\bibitem[{Utomo {et~al.}(2019)Utomo, Blitz, \& Falgarone}]{Utomo_2019}
Utomo, D., Blitz, L., \& Falgarone, E. 2019, The Astrophysical Journal, 871, 17

\bibitem[{Watkinson {et~al.}(2017)Watkinson, Majumdar, Pritchard, \&
  Mondal}]{Watkinson2017AFE}
Watkinson, C., Majumdar, S., Pritchard, J., \& Mondal, R. 2017, Monthly Notices
  of the Royal Astronomical Society, 472, 2436

\bibitem[{{Xu} {et~al.}(2016){Xu}, {Yan}, \& {Lazarian}}]{Xu2016b}
{Xu}, S., {Yan}, H., \& {Lazarian}, A. 2016, \apj, 826, 166

\bibitem[{{Yoon} \& {Cho}(2019)}]{Yoon2019ApJ...880..137Y}
{Yoon}, H. \& {Cho}, J. 2019, \apj, 880, 137

\end{thebibliography}

\appendix

\section{A. A Direct Bispectrum Estimator}
\label{appendix:estimator}

When computing the bispectrum of three-dimensional data, it is often too computationally intensive to solve for the bispectrum and bicoherence using Eqs. \ref{eq:bispectra} and \ref{eq:bicoherence}. Instead, for a given ($k_1$, $k_2$), one can do so by uniformly sampling possible combinations of vectors ($\mathbf{k}_1$, $\mathbf{k}_2$) from the sample space $\Omega$. In order to converge to $B(k_1, k_2, \theta)$, we first need to converge to the expectation value of $\tilde{\rho}(\mathbf{k}_{1}) \tilde{\rho}(\mathbf{k}_{2}) \tilde{\rho}^{*}(\mathbf{k}_1 + \mathbf{k}_2)$ by calculating a moving average. Once estimating this value, we can recover $B(k_1, k_2, \theta)$--which we defined as the sum over triangles rather than the average over triangles--by multiplying by $|\Omega|$. Note that $|\Omega|$ could be used interchangeably with a volume element. We define the estimator
\begin{equation}
    \label{eq:bispec_estimator}
    \hat{B}(k_1, k_2, \theta) = \frac{|\Omega|}{N}\sum\limits_{n = 1}^{N} 
    \tilde{\rho}(\mathbf{k}_{1}^n) \tilde{\rho}(\mathbf{k}_{2}^n) \tilde{\rho}^{*}(\mathbf{k}_1^n + \mathbf{k}_2^n)
\end{equation}
where $N$ is the number of samples and ($\mathbf{k}_{1}^{n}$, $\mathbf{k}_{2}^{n}$) are independent draws from a joint uniform distribution on $\Omega$. This is just the Monte Carlo integration of a multidimensional integral over $\Omega$.
As is standard with such an approach, $\hat{B}$ is an unbiased estimator of $B$ and the error should decrease as $1/\sqrt{N}$. More specifically, the error is the mean squared error of the quantity $\tilde{\rho}(\mathbf{k}_{1}) \tilde{\rho}(\mathbf{k}_{2}) \tilde{\rho}^{*}(\mathbf{k}_1 + \mathbf{k}_2)$ multiplied by $|\Omega|$.
We demonstrate convergence numerically in Fig. \ref{fig:Convergence}.

One can define a very similar estimator for the bicoherence
\begin{equation}
    \label{eq:bicoh_estimator}
	\hat{b}(k_1, k_2, \theta) = \frac{
		|\sum\limits_{n = 1}^N
		\tilde{\rho}(\mathbf{k}_{1}^n) \tilde{\rho}(\mathbf{k}_{2}^n) \tilde{\rho}^*(\mathbf{k}_1^n + \mathbf{k}_2^n)|}
		{\sum\limits_{n = 1}^N
		|\tilde{\rho}(\mathbf{k}_{1}^n) \tilde{\rho}(\mathbf{k}_{2}^n) \tilde{\rho}^*(\mathbf{k}_{1}^n + \mathbf{k}_{2}^n)|}.
\end{equation}
For simplicity, we use $\hat{B}$ interchangeably with $B$ and $\hat{b}$ with $b$ in this work. For our calculations on data with $N_{\rm res} = 1024^3$, we use $N = 5 \times 10^7$ samples. For $N_{\rm res} = 1024^2$, we sum over $\Omega$ exactly.

\section{B. Numerical Convergence}
\label{appendix:convergence}

 \begin{figure*}
    \includegraphics[width=\textwidth]{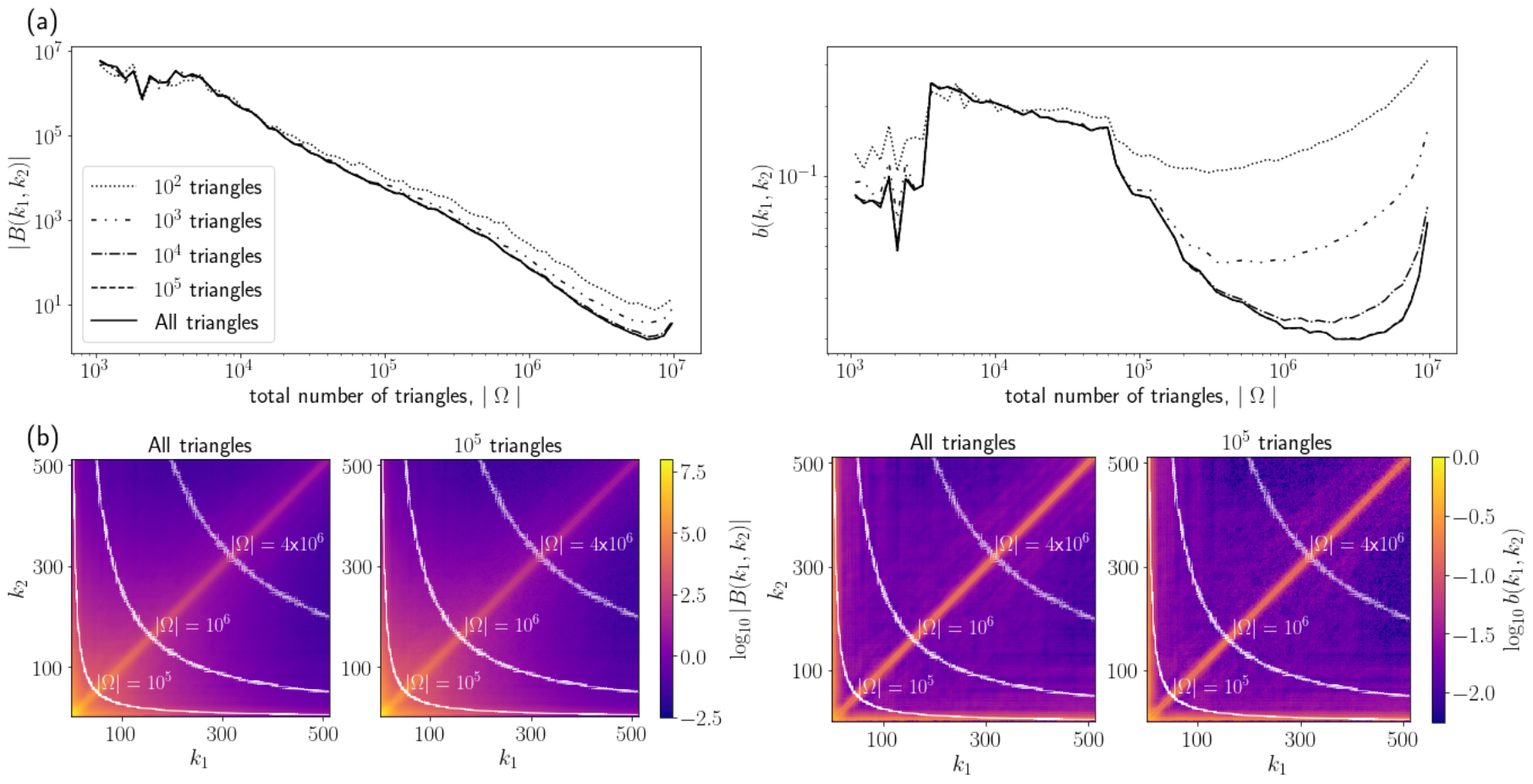}
    \caption{(a) Numerical convergence for 512x512 correlation maps of bispectral amplitudes $|B(k_1, k_2)|$ and bicoherence $b(k_1, k_2)$ calculated from $N_{\rm res} = 1024^2$ column density data with varying number of random samples $N$. We plot the profiles against $|\Omega|$, the number of available triangles to sample from. Convergence should require a larger $N$ with increasing $|\Omega|$. Convergence over all ($k_1$, $k_2$) appears to be at $N = 10^5$ samples. (b) A comparison between exact calculations of the bispectrum and bicoherence and sampling with $N = 10^5$. Contours of $|\Omega|$ are drawn in white to indicate the regions where $N = 10^5$ under-samples vs. over-samples the total number of triangles. The bicoherence exhibits more statistical noise than the bispectrum in the sampling case since it spans many less orders of magnitude, but the noise disappears with a larger number of samples.}
    \label{fig:Convergence}
\end{figure*}

Figure \ref{fig:Convergence} demonstrates the convergence of sampled Eqs. \ref{eq:bispec_estimator} and \ref{eq:bicoh_estimator}
 to exact Eqs. \ref{eq:bispectra} and \ref{eq:bicoherence}, respectively.
 Figure \ref{fig:Convergence}(a) shows the bispectral amplitudes and bicoherence binned by the size of the sample space $|\Omega|$ for an increasing number of samples $N$. For the case of an integrated column density $1024^2$ dataset, one can take the number of samples to be $10^5$ and converge to a strong estimate even for the largest $|\Omega| = 10^7$ (1\% of the total sample space). Due to the external magnetic field $\mathbf{B}_0$, our data has structural anisotropy. For data that is isotropic, we should see even faster convergence due to a smaller mean squared error.
 
 Figure \ref{fig:Convergence}(b) illustrates the bispectral amplitudes and the bicoherence for the case of exact calculations and calculation with $10^5$ samples per ($k_1$, $k_2$). The plotted contours show lines of constant $|\Omega|$ and therefore demonstrate regions where $10^5$ is considered under-sampling or over-sampling. Statistical noise due to sampling is much easier observed in the bicoherence, where the values span many less orders of magnitude than the bispectrum. This noise disappears at a larger number of samples.
 
\section{C. Implementation}
\label{appendix:implementation}
 
 We have written CPU and GPU Python implementations of Eqs. \ref{eq:bispectra}, \ref{eq:bicoherence}, \ref{eq:bispec_estimator}, and \ref{eq:bicoh_estimator} that have reduced the computational bottleneck for bispectral analysis and allow for calculations on higher resolution data compared to previous studies.
For our GPU code we use CuPy, which provides a convenient just-in-time Python interface to the CUDA programming language. We also provide an almost identically implemented CPU code using the popular package Numba.

The bispectrum implementation enumerates all points in an FFT by wavenumber $k$, and assigns an index to every possible triangle with side lengths $(k_1, \ k_2)$. This can be done without explicit calculation since the space of all possible triangles is nothing but the cartesian product of spherical shells with radii $k_1$ and $k_2$.
To calculate the bispectrum and bicoherence, the code loops over unique $(k_1, \ k_2)$ and, in parallel, gathers triangle samples, and either sums them together or bins by opening angle $\theta$. The code provides many configurable settings, which are described in its documentation.

The code is implemented in an easy-to-use python package \texttt{spatialstats}, available at https://github.com/mjo22/spatialstats. The package also includes a GPU power spectrum code and other routines for calculating correlation functions. Each routine can be used independently of the whole package and can be downloaded separately. The routines are well-documented and should be simple to use.

\section{D. Considerations with the Nyquist Frequency}
\label{appendix:corrections}

It is important to note that one must be careful when interpreting an angularly averaged bispectrum for $k_1 + k_2 > k_{\rm nyq}$, where $k_{\rm nyq}$ is the Nyquist frequency.
For these ($k_1$, $k_2$) there exist ($\mathbf{k}_1$, $\mathbf{k}_2$) in the sampling space such that $\mathbf{k}_1 + \mathbf{k}_2$ is outside of the domain of the FFT. More specifically, one cannot characterize a number of triangles such that the angle between ($\mathbf{k}_1$, $\mathbf{k}_2$) is between $0$ and $\pi/2$. Therefore, in this work the bispectrum is biased toward angles between $\pi/2$ and $\pi$ at these upper-triangular wavenumbers. The signals we consider in this paper are not in this upper triangular region.

In this region, the bispectrum $B(k_1, k_2)$ will incur some error. When we compute the bispectrum, there is good agreement with trends from the $k_1 + k_2 \leq k_{\rm nyq}$ lower-triangular region. We do not attempt an error analysis in this work, but we can revise Eq. \ref{eq:bispectra} for the angularly averaged case to mitigate error.

Under the assumption that one can still attain a good estimate for the average $\tilde{\rho}(\mathbf{k}_{1}) \tilde{\rho}(\mathbf{k}_{2}) \tilde{\rho}^{*}(\mathbf{k}_1 + \mathbf{k}_2)$ at these ($k_1$, $k_2$), we can revise the angularly averaged version of Eq. \ref{eq:bispectra} to
\begin{equation}
	\label{eq:bispec_revised}
     	B(k_{1}, k_{2}) = \frac{|\Omega|}{|\Omega'|} \sum\limits_{\Omega'}
	    	\tilde{\rho}(\mathbf{k}_{1}) \tilde{\rho}(\mathbf{k}_{2}) \tilde{\rho}^{*}(\mathbf{k}_1 + \mathbf{k}_2),
\end{equation}
where $\Omega = \Omega(k_1, k_2)$ is the usual sample space of triangles with constant side lengths and $\Omega' = \Omega'(k_1, k_2, k_{\rm nyq})$ is the sample space restricted by the Nyquist frequency.
Multiplying by the correction factor $|\Omega|/|\Omega'|$ can be thought of as first computing an average bispectrum by dividing by $|\Omega'|$ and then multiplying by the true number of triangles $|\Omega|$ that we would have access to were we not restricted by the domain of the FFT.
$|\Omega|$ is easy to compute because $|\Omega(k_1, k_2)| = |\Omega(k_1)||\Omega(k_2)|$, where $\Omega(k)$ is the number of modes in a spherical shell of radius $k$.
This is similar to the common practice of dividing by a number of modes and multiplying by a volume element.

Note that Eq. \ref{eq:bispec_revised} simplifies to Eq. \ref{eq:bispectra} for $k_1 + k_2 \leq k_{\rm nyq}$. We need not revise the bicoherence in Eq. \ref{eq:bicoherence}, besides that in reality we are sampling from $\Omega'$.

\section{E. Comparison with Bispectrum FFT Estimators}
\label{appendix:fftestimator}

 Recently, the bispectrum community in cosmology has began to utilize FFT estimators with improved computational complexity over direct triangle sampling \citep{Watkinson2017AFE, Scoccimarro2015FastEF, Regan2017Estimators}. The method can be summarized as follows.
 
 Evaluating the bispectrum for $\mathbf{k}_1 + \mathbf{k}_2 + \mathbf{k}_3 = 0$,
 
 \begin{equation}
     B(k_1, k_2, k_3) = \frac{1}{V(k_1, k_2, k_3)} \int_{k_1} d\mathbf{k}_1 \int_{k_2} d\mathbf{k}_2 \int_{k_3} d\mathbf{k}_3  \ \tilde{\rho}(\mathbf{k}_1)\tilde{\rho}(\mathbf{k}_3)\tilde{\rho}(\mathbf{k}_3) \delta(\mathbf{k}_1 + \mathbf{k}_2 + \mathbf{k}_3)
 \end{equation}
 where $\delta(\mathbf{k})$ is a Dirac-delta function, $V(k_1, k_2, k_3)$ is the integration volume, and the integral $\int_{k_j} d\mathbf{k}_j$ is over a spherical shell of wave vectors $\mathbf{k}_j$ with length $k_j$. Expanding with the definition $\delta(\mathbf{k}) = \int d\mathbf{r} \exp(i \mathbf{k}\cdot\mathbf{r})$ and writing $\exp(i \sum\limits_{j = 1}^{3}\mathbf{k}_j\cdot\mathbf{r}) = \prod\limits_{j = 1}^{3}\exp(\mathbf{k}_j\cdot\mathbf{r})$,
 
  \begin{equation}
     B(k_1, k_2, k_3) = \frac{1}{V(k_1, k_2, k_3)} \int d\mathbf{r} \prod\limits_{j = 1}^{3} \int_{k_j} d\mathbf{k}_j \ \tilde{\rho}(\mathbf{k}_j) \exp(i \mathbf{k}_j\cdot\mathbf{r}).
 \end{equation}
 We can obtain the three $\int_{k_j} d\mathbf{k}_j \ \tilde{\rho}(\mathbf{k}_j) \exp(i \mathbf{k}_j\cdot\mathbf{r})$ terms by computing an inverse FFT of $\tilde{\rho}(\mathbf{k}_j)$ multiplied by a binning function $a(\mathbf{k}_j)$ which is $1$ for all $|\mathbf{k}_j| = k_j$ and $0$ otherwise. Then the integral $\int d\mathbf{r}$ is computed as a sum over all grid points $\mathbf{r}$. This definition for the FFT estimator is
 
 \begin{equation}
     \hat{B}(k_1, k_2, k_3) = \frac{1}{V(k_1, k_2, k_3)} \sum\limits_{\mathbf{r}} \prod\limits_{j = 1}^{3} \text{IFFT} [\tilde{\rho}(\mathbf{k}_j) a(\mathbf{k}_j)].
 \end{equation}
 Standard methods can be used to calculate $V(k_1, k_2, k_3)$.
 
 Our Monte Carlo approach to sampling may have practical advantages in some use cases.
 It is parallelized over a collection of triangles and loops over $(k_1, \ k_2)$ to create maps of wavenumbers over a range of $\theta$. The primary memory requirements are one FFT and the $N$ terms of the sum in Eq. \ref{eq:bispec_estimator}.
 Direct sampling is less memory intensive than FFT estimators and one can fit larger datasets on a GPU.
 In this work we calculate bispectra for $N_{\rm res} = 1024^3$ in double precision on a Nvidia A100 40GB GPU, which would not be possible if we needed to store 6 FFTs in memory as indicated in \cite{Watkinson2017AFE}.
 In our use case, a true speed comparison pits a GPU Monte Carlo implementation against a CPU FFT estimator method.
 
 It is also a bit non-trivial to calculate the bispectrum  over a large range of $(k_1, \ k_2, \ k_3)$ with the FFT estimator method. If one wants to sweep over all wavenumbers, ideally the FFT estimator method could recycle many of the 6 FFTs per $(k_1, \ k_2, \ k_3)$. However, CPU memory limitations for large datasets may require a user to recompute or store some of these in memory. While this is still often much faster than measuring all possible triangles, if it is appropriate to measure only a subset of them then our GPU Monte Carlo code can sweep over many $(k_1, \ k_2, \ k_3)$ without added bottlenecks.
 
 Another key advantage is that the Monte Carlo code is parallel for a given $(k_1, \ k_2)$--\textit{not} for a $(k_1, \ k_2, \ \theta)$. The FFT estimator adds another sequential loop since it is parallel per $(k_1, \ k_2, \ k_3)$. This is another reason why our implementation is preferable when computing grid points in bulk, like the 2D bispectral correlation maps that we have presented in this paper.
 
 Use cases for the bispectrum where one can relax the number of samples $N$ are particularly appealing for Monte Carlo.
 In the case of perfectly isotropic data, we expect rapid convergence with increasing $N$ since the variance over all triangle evaluations should be quite small.
 This tunable complexity is also advantageous if a user is more interested in qualitative features of the bispectrum opposed to measuring precise amplitude scaling. They can relax the computation and get an answer quickly.
 
 Recent developments in bispectrum FFT estimators have clear advantages in computational complexity. They make exact computation feasible when counting all triangles is not. However, they are not always optimal. For some cases, a GPU Monte Carlo algorithm could prove to be faster and simpler to use.

\end{document}